\documentclass[sigconf]{acmart}

\usepackage{times}  
\usepackage{helvet}  
\usepackage{courier}  

\usepackage{graphicx} 
\usepackage{natbib}  
\usepackage{caption,subcaption} 
\usepackage{algorithm}
\usepackage{algorithmic}

\usepackage{booktabs}
\usepackage{multirow}
\usepackage{longtable}

\usepackage{newfloat}
\usepackage{listings}
\acmConference[HEAL@CHI'25]{2nd HEAL Workshop at CHI Conference on Human Factors in Computing Systems}{April 26, 2025}{Yokohama, Japan}

\DeclareCaptionStyle{ruled}{labelfont=normalfont,labelsep=colon,strut=off} 
\floatstyle{ruled}
\newfloat{listing}{tb}{lst}{}
\floatname{listing}{Listing}
\title{EvalAssist: A Human-Centered Tool for LLM-as-a-Judge}

\author{Zahra Ashktorab}
\email{zahra.ashktorab@ibm.com}
\affiliation{%
  \institution{IBM Research}
  \city{Yorktown Heights}
  \state{NY}
  \country{USA}
}

\author{Werner Geyer}
\email{werner.geyer@ibm.com}
\affiliation{%
  \institution{IBM Research}
  \city{Cambridge}
  \state{MA}
  \country{USA}
}

\author{Michael Desmond}
\email{michael.desmond@ibm.com}
\affiliation{%
  \institution{IBM Research}
  \city{Yorktown Heights}
  \state{NY}
  \country{USA}
}

\author{Elizabeth M. Daly}
\email{elizabeth.daly@ibm.com}
\affiliation{%
  \institution{IBM Research}
  \city{Dublin}
  \country{Ireland}
}

\author{Martín Santillán Cooper}
\email{martin.cooper@ibm.com}
\affiliation{%
  \institution{IBM Research}
  \city{Yorktown Heights}
  \state{NY}
  \country{USA}
}

\author{Qian Pan}
\email{qian.pan@ibm.com}
\affiliation{%
  \institution{IBM Research}
  \city{Cambridge}
  \state{MA}
  \country{USA}
}

\author{Erik Miehling}
\email{erik.miehling@ibm.com}
\affiliation{%
  \institution{IBM Research}
  \city{Dublin}
  \country{Ireland}
}

\author{Tejaswini Pedapati}
\email{tejaswini.pedapati@ibm.com}
\affiliation{%
  \institution{IBM Research}
  \city{Yorktown Heights}
  \state{New York}
  \country{USA}
}

\author{Hyo Jin Do}
\email{hjdo@ibm.com}
\affiliation{%
  \institution{IBM Research}
  \city{Cambridge}
  \state{MA}
  \country{USA}
}

\begin{document}

\begin{abstract}
With the broad availability of large language models and their ability to generate vast outputs using varied prompts and configurations, determining the best output for a given task requires an intensive evaluation process—one where machine learning practitioners must decide how to assess the outputs and then carefully carry out the evaluation. This process is both time-consuming and costly. As practitioners work with an increasing number of models, they must now evaluate outputs to determine which model and prompt performs best for a given task. LLMs are increasingly used as evaluators to filter training data, evaluate model performance, assess harms and risks, or assist human evaluators with detailed assessments. We present \emph{EvalAssist}, a framework that simplifies the LLM-as-a-judge workflow. The system provides an online criteria development environment, where users can interactively build, test, and share custom evaluation criteria in a structured and portable format. We support a set of LLM-based evaluation pipelines that leverage off-the-shelf LLMs and use a prompt-chaining approach we developed and contributed to the UNITXT open-source library. Additionally, our system also includes specially trained evaluators to detect harms and risks in LLM outputs. We have  deployed the system internally in our organization with several hundreds of users.
\end{abstract}

\settopmatter{printacmref=false} 
\renewcommand\footnotetextcopyrightpermission[1]{%
  \footnotetext{%
    \textit{2nd HEAL Workshop at CHI Conference on Human Factors in Computing Systems, April 26, 2025, Yokohoma Japan. Please cite the published version.} \\
    \copyright~2025 Copyright held by the owner/author(s).
  }
}
\maketitle



%

\section{Keywords}
large language models, llm-as-a-judge, evaluation tools, direct assessment, pairwise comparison
\section{Introduction}

Human evaluation of Large Language Models (LLMs) is common practice and still considered gold standard. However, given cost and time constraints, LLMs are increasingly being used to judge the output of other LLMs, often referred to as LLM-as-a-judge. This approach is attractive as it can accommodate use case specific needs through custom criteria, is easy-to-understand by non-technical users, does not require reference data, and can significantly reduce human evaluation effort. Empirical studies have reported high agreement between LLM and human ratings. For example \cite{zheng2023large} report more than 80\% agreement and \cite{kim2023prometheus} report that fine tuned evaluator models have high correlation with human judgments. More recently, evaluator ensembles have been shown to be effective \cite{verga2024replacing}.

While LLM-as-a-judge has become popular, several challenges remain to make the approach effective, trustworthy and aligned with user needs. Doddapaneni et al. \citep{doddapaneni2024finding} report on evaluator LLMs that failed to identify synthetic quality drops in half the cases, suggesting that evaluators did not understand the task. Bavaresco et al. \citep{bavaresco2024llms} urge caution after empirical analysis of several language tasks and conclude that LLMs are not yet ready to systematically replace human judges. Finally, issues with bias \cite{liu2303g, li2023alpacaeval} and prompt sensitivity \cite{wang2023chatgpt} have been identified that prevent the straightforward, out-of-the-box use of LLMs as judges. In other words, one cannot simply deploy an LLM to evaluate content reliably; instead, additional validations and safeguards are required for practical application.

In this demo, we introduce \texttt{EvalAssist}, a comprehensive LLM-as-a-judge framework that provides a criteria development and test environment, in which users can iteratively test and refine evaluation criteria until they are confident that they work well and align with their expectations. The criteria can then be applied to a larger dataset by exporting a Jupyter notebook from EvalAssist which is based on the UNITXT open-source evaluation library \cite{bandel2024unitxt}. Our algorithms available in UNITXT are inspired by Chain-of-Thought prompting and  and we compute positional bias as a metric for uncertainty, which can help engender trust in the evaluation process. We also plan to open source the EvalAssist UI framework built on top of UNITXT.

\section{Design Goals}
In designing \texttt{EvalAssist}, we aim to address the challenges that engineers, data scientists, and researchers face when evaluating outputs and aligning criteria with their requirements. We identify five design goals to guide the development of \texttt{EvalAssist}, addressing key gaps in existing tools. We build on user findings from prior work with ML practitioners, which highlights the challenges faced during model evaluation \cite{pan2024human,desmond2024evalullm}. Our approach is designed to support engineers, ML practitioners, and researchers by incorporating key features such as the ability to choose between pairwise comparison and direct assessment, (positional) bias indicators, and scalable, cost-efficient workflows.  Below, we outline these goals and how \texttt{EvalAssist} stands apart in the evaluation tool landscape:
\\

\textbf{DG1. Isolate Generation from Evaluation}. Isolating generation and prompt engineering from evaluation is a significant need for engineers, data scientists, and practitioners working in complex multi-agent environments or retrieval-augmented generation (RAG) contexts \cite{gao2023retrieval}, where sophisticated workflows often exist outside the evaluation tool and result in large datasets across various models, prompts, and configurations \cite{pan2024human,desmond2024evalullm}. Other tools couple the prompt engineering process with the evaluation process \cite{shankar2024validates, arawjo2024chainforge, kim2023evallm} as they focus on prompting and variations that result in output as a result of varied prompts. The targeted users for \texttt{EvalAssist} are developers, including engineers, ML practitioners, and researchers, who often have access to a wider variety of models and need to design complex LLM workflows to select the best model for a specific task. 

\textbf{DG2. Reduce Cost of Evaluation.} Applying AI-assisted evaluation to large datasets typically used by research engineers, data scientists \cite{ZhengChiangSheng2024} is costly and time consuming. Prior work has shown that users would like to run an evaluation on a subset of the data first \cite{pan2024human}. \texttt{EvalAssist} addresses the challenge of costly and time-consuming model calls by allowing users to evaluate a subset of the data, refine their criteria, and then, once satisfied, apply these criteria to the entire dataset through an Software Development Kit (SDK)— a crucial feature for users handling very large datasets.

\textbf{DG3. Support Multiple AI Evaluators.} Users can select their preferred model from a set of LLMs as the evaluator, addressing issues like self-enhancement bias \cite{xu2024pride}, where models favor their own responses. EvalAssist is model-agnostic, enabling users to choose any model for evaluation, unlike other front-end tools that limit users to a single AI evaluator.

\textbf{DG4. Include Bias Indicators.} ML researchers using LLM-as-a-judge are aware of potential biases exhibited by AI evaluation and have expressed the desire to have transparent indicators on bias in LLM-as-a-judge tooling \cite{pan2024human}. Positional bias occurs when a model consistently favors one option based solely on its position, rather than its content \cite{li2024split}. In evaluation tasks, such as pairwise comparisons, this bias arises if the model disproportionately selects the output to be evaluated in a particular position (e.g., always favoring the first output) regardless of the actual content. In direct assessments, where a response is evaluated based on specific criteria and options, the order of these options can be shuffled. \texttt{EvalAssist} is the only LLM-as-a-judge application that provides a positional bias indicator in the user experience. Including bias indicators helps users recognize positional uncertainty in LLM judgments. 

\textbf{DG5. Enable Flexible Evaluation Methods} \texttt{EvalAssist} is the first tool to allow users to define their criteria for the two most common LLM-as-a-judge approaches: direct assessment (score-based, single-grading) and pairwise comparison (relation-based) \cite{ZhengChiangSheng2024,shi2024judging}. With this tool, users can select the strategy that best suits their dataset or explore both approaches to determine which is more effective for their needs.

\section{EvalAssist: System Design} 

\begin{figure*}
    \centering
 \includegraphics[width=.9\linewidth]{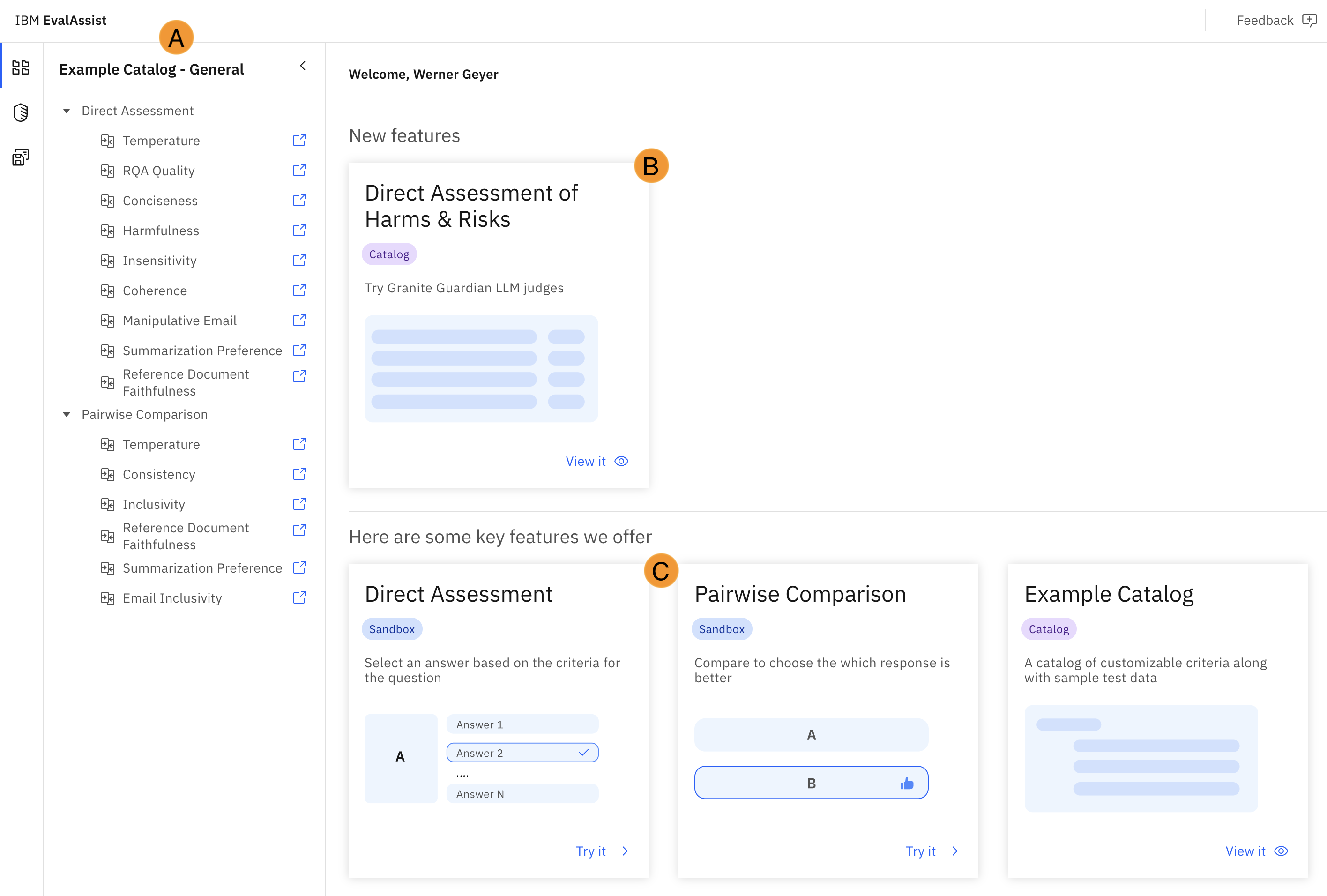}
    \caption{EvalAssist landing page with test case catalog on the left and different evaluation strategies to choose from in the center.}
    \Description{Screenshot of EvalAssist landing page showing a test case catalog on the left and evaluation strategies in the center.}
    \label{fig:EvalAssist Overview}
\end{figure*}

\texttt{EvalAssist} abstracts the LLM-as-a-Judge evaluation process into parameterize-able evaluators (the criterion being the parameter), allowing the user
to focus on criteria definition. EvalAssist is independent of the LLM used to generate the output to be evaluated, i.e. this approach acknowledges that developers often use complex external workflows to adjust configurations (e.g., model temperature) and experiment with different models and prompts to generate responses \cite{desmond2024evalullm}.  \texttt{EvalAssist} consists of a web-based user experience (see Figure \ref{fig:EvalAssist Overview}) and an API based on the UNITXT open-source evaluation library \cite{bandel2024unitxt}. The user interface provides a convenient way of iteratively testing and refining LLM-as-a-judge criteria, and supports both direct (rubric-based) and pairwise assessment paradigms (relation-based), the two most prevalent forms of LLM-as-a-judge evaluation available \cite{kim2023prometheus,ZhengChiangSheng2024}. 
\texttt{EvalAssist} can leverage both off-the-shelf instruction-tuned models to be used as  evaluators or specialized judge models such as, for example, Granite Guardian \cite{padhi2024graniteguardian}, a judge model to assess harms and risks (see \ref{fig:EvalAssist Overview} B). Since \texttt{EvalAssist} uses UNITXT as it's main judging API, new models can be easily incorporated by adding them through UNITXT. Once users are satisfied with their criteria, they can run bulk evaluations with larger data sets by downloading a Jupyter Notebook that contains their criteria definition and the necessary code to run large evaluations with UNITXT. We also allow users to save their test cases and provide a catalog of predefined criteria (see Figure \ref{fig:EvalAssist Overview} A). A test case in the Example Catalog includes a criteria definition and the data being evaluated.

\subsection{Task Context}
\begin{figure*}[htbp]
    \centering
    \includegraphics[width=0.8\linewidth]{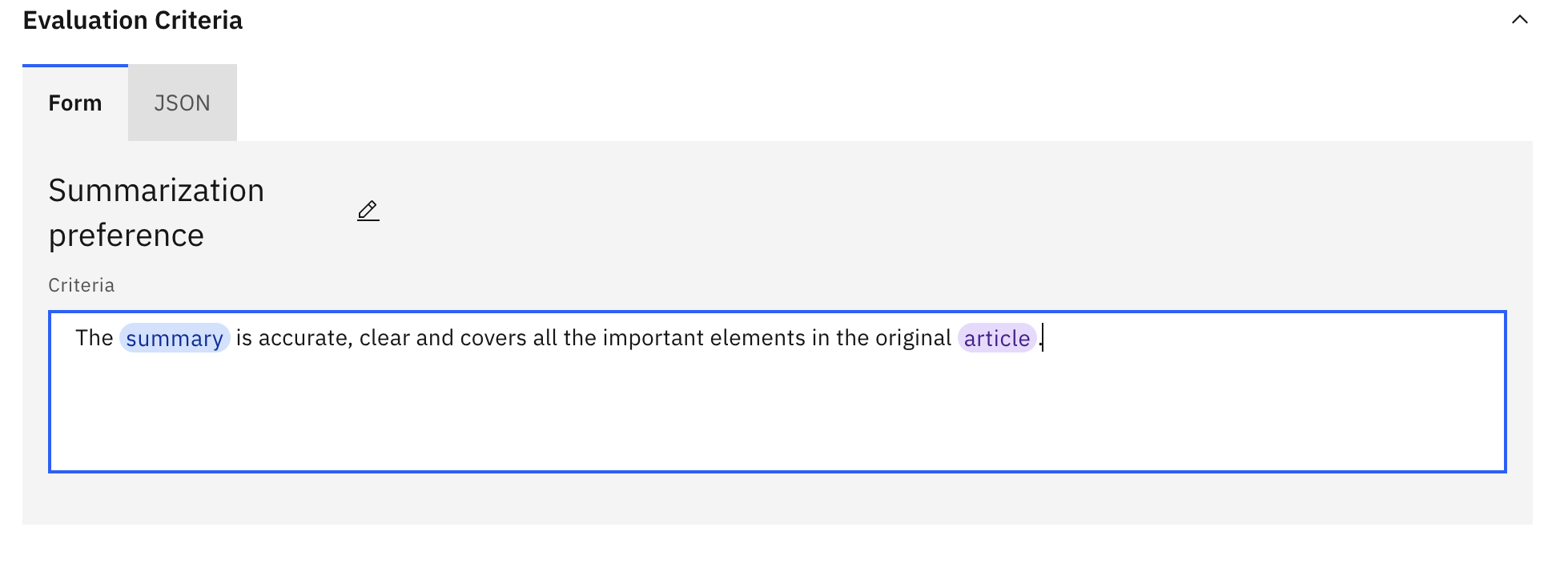}
    \caption{Evaluation Criteria Form for Pairwise Comparison. Variables created in the task context can be referenced in the criteria definition.}
    \label{fig:evaluation-criteria-pairwise}
\end{figure*}

\begin{figure*}[htbp]
    \centering
    \includegraphics[width=0.7\linewidth]{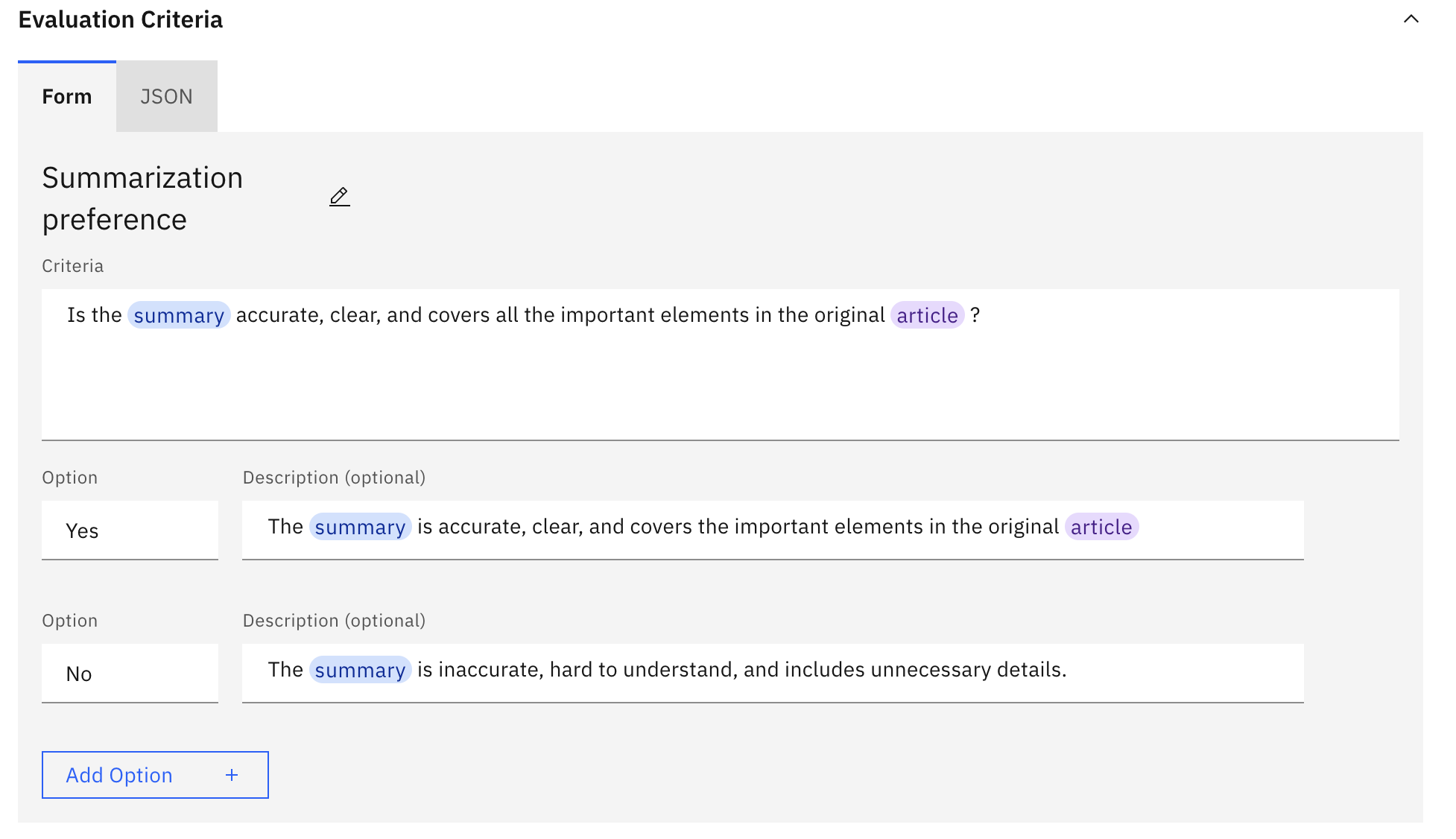}
    \caption{Evaluation Criteria Form for Direct Assessment. Variables created in the task context can be referenced in the criteria definition and in the options.}
    \label{fig:evaluation-criteria-direct}
\end{figure*}

\begin{figure*}[ht]
    \centering
 \includegraphics[width=1\linewidth]{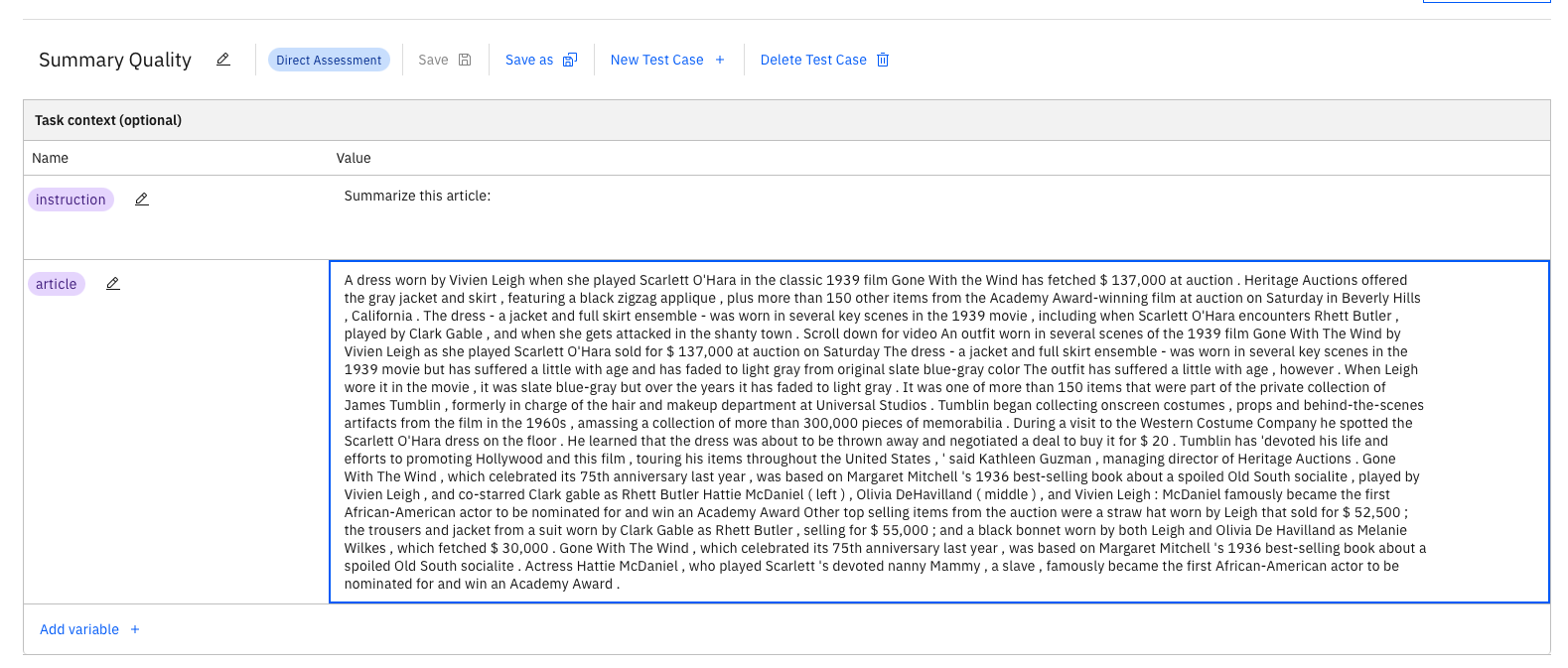}
    \caption{Task Context for a Summarization Task. The Task Context is consistent for both direct assessment and pairwise comparison strategies. Users have the option to break down the context into variables, such as the instruction and article, to simplify and unify references when developing evaluation criteria.}
    \label{fig:taskcontext}
\end{figure*}

To run LLM-as-a-judge evaluations in EvalAssist, users need to create a new test case by choosing between direct assessments and pairwise comparisons (Figure \ref{fig:EvalAssist Overview}C). Once a new test case has been created, 
users can optionally define task-relevant input data through variables in the task context (Figure \ref{fig:taskcontext}), such as, for example, the prompt, the article to summarize, or the source data for content-grounded Q\&A. Variables make it easier to reference these elements when defining criteria in the evaluation forms (Figures \ref{fig:evaluation-criteria-pairwise} and \ref{fig:evaluation-criteria-direct}).

\subsection{Direct Assessment}
\begin{figure*}
    \centering
 \includegraphics[width=.9\linewidth]{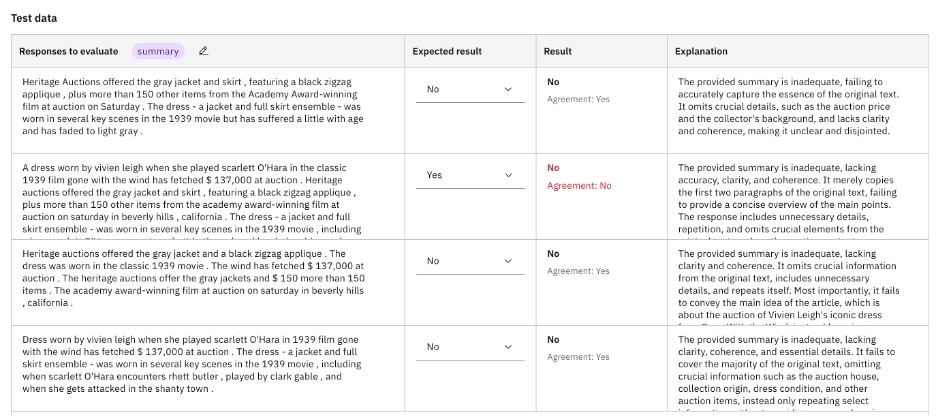}
    \caption{Results for direct assessment. Users can select their expected judgments for the output, which are auto-populated based on the criteria they define (i.e., the scale items created when setting the criteria). The results display the AI Evaluator's judgments, indicating whether there is agreement between the user and the AI, along with explanations for each result.}
    \label{fig:rubricresults}
\end{figure*}

 With this strategy, users evaluate outputs based on a single criterion rubric they define. The evaluation criteria form (Figure \ref{fig:evaluation-criteria-direct}) allows defining criteria with a title, criteria description, and an arbitrary number of free-form options. These are the options the LLM Evaluator will have to choose from during assessment. As such, the 
 system supports both binary and multi-level scale assessments. In the Evaluator section (not shown), users must select one of the pre-configured LLM evaluators.  In the Test Data Section (see Figure \ref{fig:rubricresults}), users enter the outputs they want to evaluate. 
 
 
 Additionally, users can optionally enter the result they would expect to see for each output. This feature is useful, if users evaluate a large number of items and want to see at a glance which evaluations failed. After running the evaluation, the system shows the actual results next to the expected results, including agreement, positional bias if present, and an explanation.

\begin{figure*}[h!]
    \centering
    \begin{subfigure}[b]{0.68\textwidth}
        \centering
        \includegraphics[width=\textwidth]{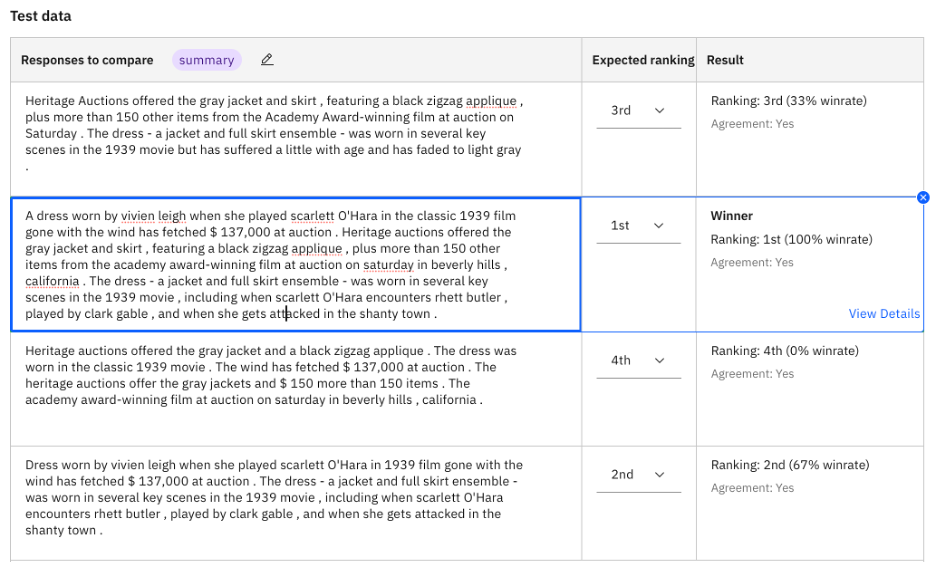}
        \caption{Ranking results generated from pairwise comparison assessment. Users can input their expected ranking to and assess their level of agreement with the AI evaluator.}
        \label{fig:ranking2}
    \end{subfigure}
  \hspace{0.01\textwidth}
    \begin{subfigure}[b]{0.28\textwidth}
        \centering
       \includegraphics[width=\textwidth]{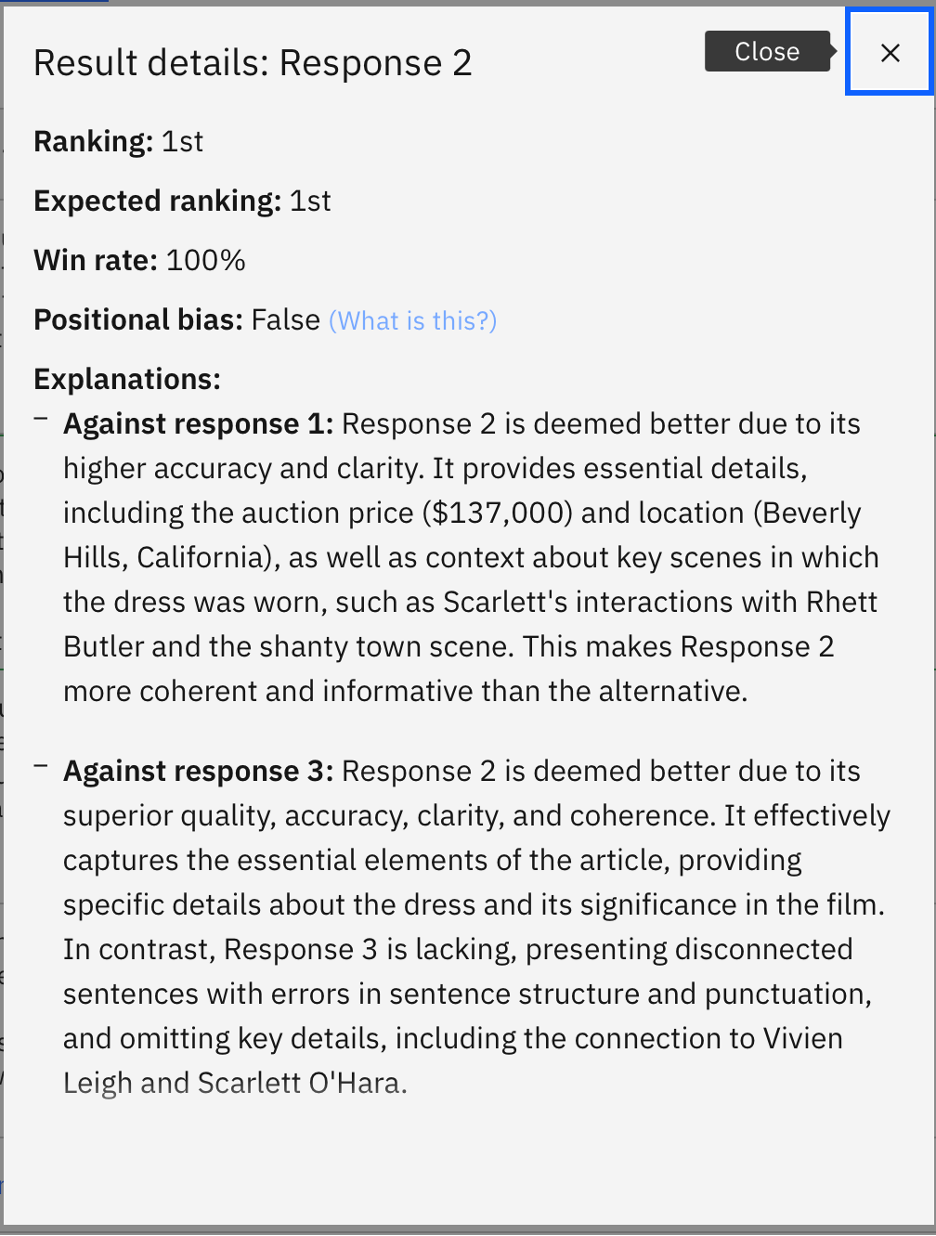}
        \caption{Explanations for each pairwise comparison in pairwise assessment.}
        \label{fig:figure2b}
    \end{subfigure}
    \caption{Results, explanations, and expected ranking generated through pairwise comparison. }
    \label{fig:sidebyside2}
\end{figure*}

\begin{figure*}[h]
    \centering
 \includegraphics[width=.8\linewidth]{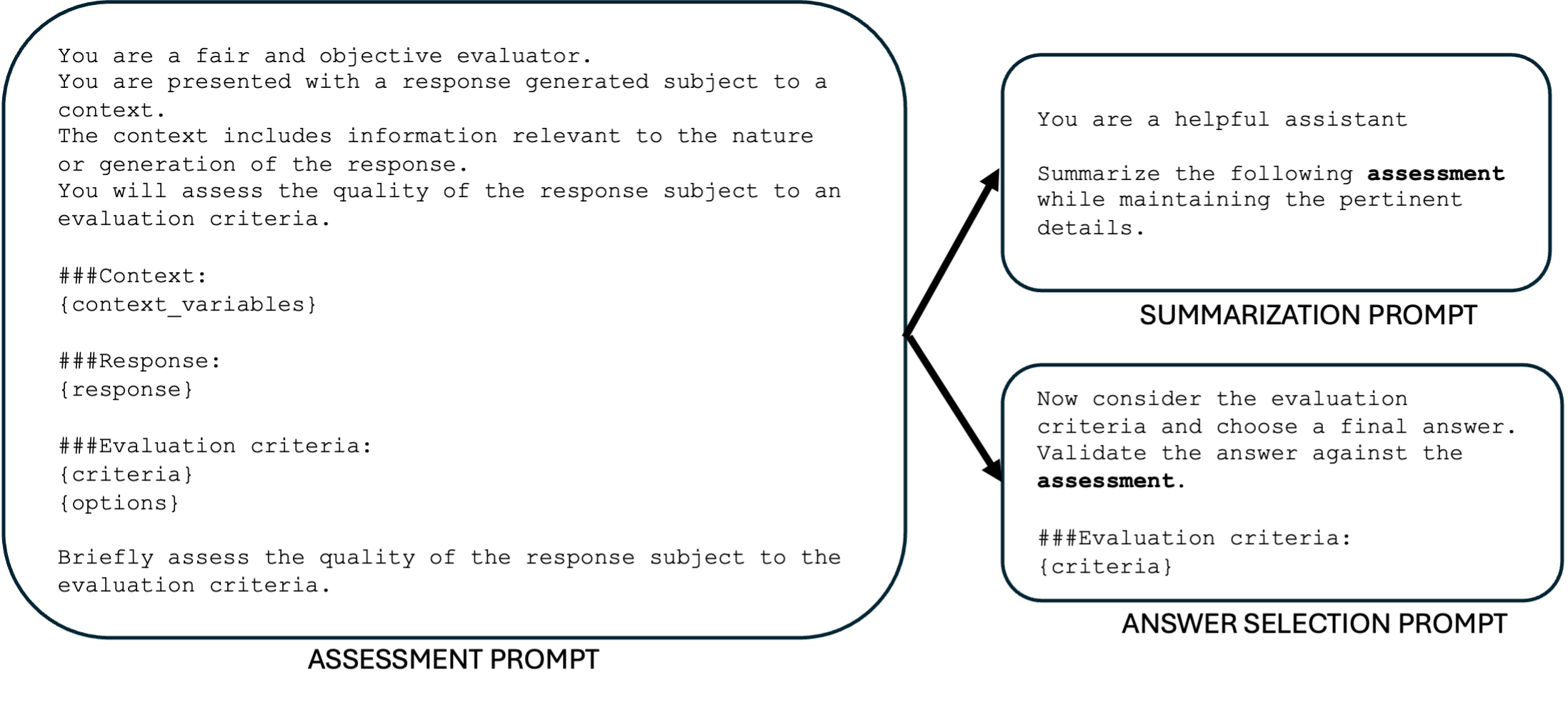}
    \caption{The assessment prompt, summarization prompt and answer selection prompt for one of the AI evaluators in the direct assessment context. The assessment prompt is used first, generating an evaluation of the response based on context and evaluation criteria. This assessment is then passed to a summarization prompt, which condenses the findings, and an answer selection prompt, which validates the response against the evaluation criteria.}
    \label{fig:prompts}
\end{figure*}
\subsection{Pairwise Comparison}
In this strategy, \texttt{EvalAssist} compares two or more outputs pairwise against one another selecting the one that better fits the criteria. The best output is determined by computing the win rate across all pairwise output comparisons. Similar to direct assessment, users can provide task-relevant input data through variables, define a criteria, and select an evaluator LLM. However, options don't need to be added to pairwise comparisons (Figure \ref{fig:evaluation-criteria-pairwise}). After evaluation, we display the results next to the expected results (see Figure \ref{fig:ranking2}), including the winner, ranking, and agreement with expected ranking. Explanations in pairwise comparison can be seen in Figure \ref{fig:figure2b} and are generated as a result of comparing pairs of responses to be evaluated. In total, $\binom{N}{2}$ comparisons are performed in a pairwise manner, where N is the total number of outputs being evaluated. Each pairwise comparison generates an explanation. The outputs are then ranked based on a win rate metric, similar to \cite{dubois2024alpacafarm}. The explanations presented in Figure \ref{fig:figure2b} correspond to the comparisons of each summary against the highlighted row in Figure \ref{fig:ranking2}. As a result, users are able to click on each result to see detailed explanations including positional bias, win-rate, and explanations for the comparisons with the other outputs.

\subsection{Positional Bias}
 To test for positional bias, an evaluation is conducted twice with the options to choose from or the outputs to be compared presented in different orders. If the outcomes differ between the two evaluations, positional bias is present, as it indicates that the position influenced the evaluation. Conversely, if the outcomes are the same, it suggests that the model's assessment was not influenced by the position. In \texttt{EvalAssist}, we include a positional bias indicator for both direct and pairwise assessments. This indicator is displayed for each row in the results and is flagged in red text to highlight inconsistent judgments by the AI evaluators.


\subsection{Evaluation}
When users select the "Evaluate" button, their input is sent to the chosen evaluator. Each evaluator is designed to perform either direct assessment or pairwise comparison. The main external difference between these two lies in how the input criteria is structured. Internally, evaluators operate as a dialog with the associated LLM using a set of custom prompts specific to that AI Evaluator. The process consists of an assessment prompt, a summarization prompt, and an answer selection prompt for all AI evaluators as shown in Figure \ref{fig:prompts}. However, specialized judges such as Granite Guardian use only a single prompt, as they have already been trained to provide high-quality evaluations and explanations \cite{padhi2024graniteguardian}. 

First, the LLM is prompted to review the evaluation task, considering the task context, criteria, and subject of evaluation. The LLM generates an open-ended assessment that explains its decision-making process. This step is inspired by Chain-of-Thought (CoT) prompting \cite{wei2022chain}, encouraging the LLM to base its final judgement on its initial reasoning.  This generated assessment is then added to the dialog history. Next, the LLM is asked to make a final judgement taking the assessment into account.
Finally, the summarization prompt results in a summarization of the initial assessment which serves as the explanation presented to the user.  Positional bias is checked by shuffling the order of the options presented to the LLM and verifying the consistency of its final decision. EvalAssist’s evaluation algorithms have been open-sourced as part of UNITXT, a flexible library for customizable textual data preparation and evaluation designed for generative language models \cite{bandel2024unitxt}. These capabilities are available for use at \cite{unitxt_llm_as_judge}. Additionally, we are working on open-sourcing the front end, allowing users to access its features while running AI-assisted evaluations.

\section{User Evaluation}


\texttt{EvalAssist} has been deployed internally, with over 700 users so far and we have conducted multiple controlled user studies with internal users.  
Our research \cite{ashktorab2024aligning} 
indicates that users prefer direct assessment when they seek greater control, while pairwise assessment is favored for more subjective tasks.  We also observed significant variability in how users defined subjective criteria, highlighting the challenge of aligning AI-assisted evaluations with diverse stakeholder needs. Some users over-specified their criteria, tailoring them too closely to a single example, while others provided overly vague definitions, relying on the AI evaluator to interpret key concepts. These differences emphasize the importance of clear stakeholder deliberation when defining evaluation criteria.  

While trust levels remained consistent between direct assessment and pairwise comparison, explanation visibility played a key role. Users found direct assessment explanations more useful, suggesting a need to improve how explanations are presented in pairwise evaluations. Bias indicators were also highly valued, pointing to opportunities to expand bias detection beyond positional bias. Finally, users expressed a strong preference for flexible evaluation strategies, using direct assessment for structured tasks and pairwise ranking for more subjective judgments. 

\section{Conclusion}
\texttt{EvalAssist} streamlines LLM-as-a-Judge workflows by enabling users to define, test, and refine evaluation criteria with transparency and flexibility. By supporting direct assessment and pairwise comparison, integrating multiple AI evaluators, and incorporating bias indicators, \texttt{EvalAssist} enhances trust and usability in AI-assisted evaluations. 
User studies revealed a preference for direct assessment in structured tasks and pairwise comparison for subjective ones. 





\bibliographystyle{ACM-Reference-Format}
\bibliography{main}

\newpage

\appendix

\end{document}